\documentclass[aps,prl,reprint,superscriptaddress,showpacs]{revtex4-1}

\usepackage{graphicx}
\usepackage{amsmath}
\usepackage[loose]{units}
\usepackage{textcomp}
\usepackage[T1]{fontenc}
\usepackage{amsmath}

\begin{document}
	\title{Magnetic Excitations in the Multiferroic N\'eel-type Skyrmion Host GaV$_4$S$_8$}
	\author{D.~Ehlers}
	\affiliation{Experimentalphysik V, Elektronische Korrelationen und Magnetismus, Universit\"at Augsburg, 86135 Augsburg, Germany}
	\author{I.~Stasinopoulos}
	\affiliation{Lehrstuhl f\"ur Physik funktionaler Schichtsysteme, Technische Universit\"at M\"unchen, Physik Department, 85748 M\"unchen, Germany}
	\author{V.~Tsurkan}
	\affiliation{Experimentalphysik V, Elektronische Korrelationen und Magnetismus, Universit\"at Augsburg, 86135 Augsburg, Germany}
	\affiliation{Institute of Applied Physics, Academy of Sciences of Moldova, MD 2028, Chisinau, Republica Moldova}
	\author{H.-A.~Krug von Nidda}
	\affiliation{Experimentalphysik V, Elektronische Korrelationen und Magnetismus, Universit\"at Augsburg, 86135 Augsburg, Germany}
	\author{T.~Feh\'er}
	\affiliation{Department of Physics, Budapest University of Technology and Economics and MTA-BME Lend\"ulet Magneto-optical Spectroscopy Research Group, 1111 Budapest, Hungary}
	\author{A. Leonov}
	\affiliation{IFW Dresden, Postfach 270016, 01171 Dresden, Germany}
	\author{I.~K\'ezsm\'arki}
	\affiliation{Department of Physics, Budapest University of Technology and Economics and MTA-BME Lend\"ulet Magneto-optical Spectroscopy Research Group, 1111 Budapest, Hungary}
	\author{D.~Grundler}
	\affiliation{Lehrstuhl f\"ur Physik funktionaler Schichtsysteme, Technische Universit\"at M\"unchen, Physik Department, 85748 M\"unchen, Germany}
	\affiliation{EPFL STI IMX LMGN, MXC 241, Station 12, 1015 Lausanne, Switzerland}
	\author{A.~Loidl}
	\affiliation{Experimentalphysik V, Elektronische Korrelationen und Magnetismus, Universit\"at Augsburg, 86135 Augsburg, Germany}
	\date{\today}

	\begin{abstract}
		Broadband microwave spectroscopy has been performed on single-crystalline GaV$_4$S$_8$, which exhibits a complex magnetic phase diagram including cycloidal, N\'eel-type skyrmion lattice, as well as field-polarized ferromagnetic phases below \unit[13]{K}. At zero and small magnetic fields two collective modes are found at 5 and \unit[15]{GHz}, which are characteristic of the cycloidal state in this easy-axis magnet. In finite fields, entering the skyrmion lattice phase, the spectrum transforms into a multi-mode pattern with absorption peaks near 4, 8, and \unit[15]{GHz}. The spin excitation spectra in GaV$_4$S$_8$ and their field dependencies are found to be in close relation to those observed in materials with Bloch-type skyrmions. Distinct differences arise from the strong uniaxial magnetic anisotropy of GaV$_4$S$_8$ not present in so-far known skyrmion hosts.
	\end{abstract}

	\maketitle

	The occurence of nontrivial topology in the spin pattern of magnets has gained considerable interest in condensed matter physics. Recent research focuses on magnetic skyrmions which are thermodynamically stabilized in compounds with noncentrosymmetric crystal structures, in a limited region of the magnetic field versus temperature phase diagram \cite{Bogdanov1989, Bogdanov1994a, Bogdanov1994b}. Skyrmions are whirl-like objects of spins which can crystallize in skyrmion lattices (SkLs) with typical lattice constants from ten to hundred nanometers and give rise to emergent electrodynamics, like the topological Hall effect \cite{Neubauer2009, Kanazawa2011, Ritz2013} or magnetic monopoles \cite{Milde2013}. Individual skyrmions have been proposed as building blocks for novel nanomagnetic storage devices \cite{Fert2013, Sampaio2013}. The SkL has raised high interest for microwave-technology applications after collective spin excitations predicted in the GHz range \cite{Mochizuki2012} were evidenced in the insulating chiral magnet Cu$_2$OSeO$_3$~\cite{Seki2012a, Onose2012, Okamura2013, Okamura2015, Zhang2015, Seki2015}. Later it was found that different metallic, semiconducting, and insulating chiral magnets support the same set of characteristic excitations, i.e., three SkL modes characterized as clockwise (CW), counterclockwise (CCW) and breathing (BR) modes, that all follow a universal behavior \cite{Schwarze2015}. 
	
\begin{figure}
		\includegraphics[width=7cm]{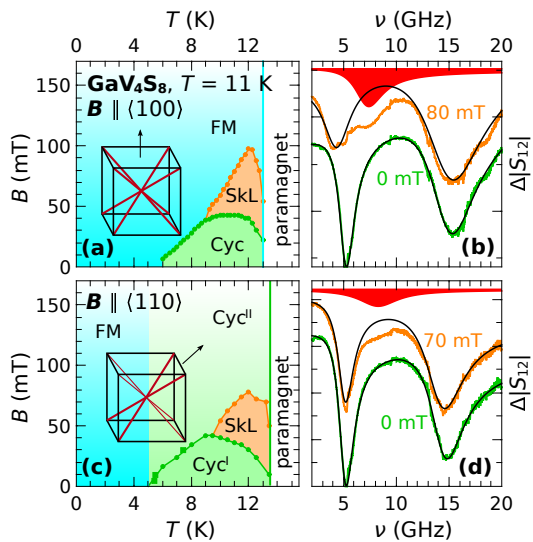}
		\caption{(Color online) (a) and (c) Magnetic phase diagrams of GaV$_4$S$_4$ for field directions $B \parallel \langle 100 \rangle$ and $B \parallel \langle 110 \rangle$, respectively, derived from detailed magnetization measurements \cite{Kezsmarki2015a}. Cycloidal (Cyc, Cyc$^\text{I}$, Cyc$^\text{II}$) and skyrmion-lattice spin structures (SkL) are embedded in a ferromagnetic (FM) phase. Cubes and arrows illustrate the orientation of external field with respect to the the possible directions of the rhombohedral distortion, i.e. the body diagonals of the cube. (b) and (d) Selected spectra at temperature $T = \unit[11]{K}$ for field directions $B \parallel \langle 100 \rangle$ and $B \parallel \langle 110 \rangle$, respectively, with decompositions of the Dysonian fits (black lines, see text) for \unit[80]{mT} and \unit[70]{mT}. The additional excitations in the SkL phase, obtained after subtraction of the two dominant Dysonian peaks, are highlighted in red. For comparison the zero-field spectra with almost perfect fits are shown.}
		\label{PDe}
	\end{figure}

\begin{figure}
		\includegraphics[width=7.5cm]{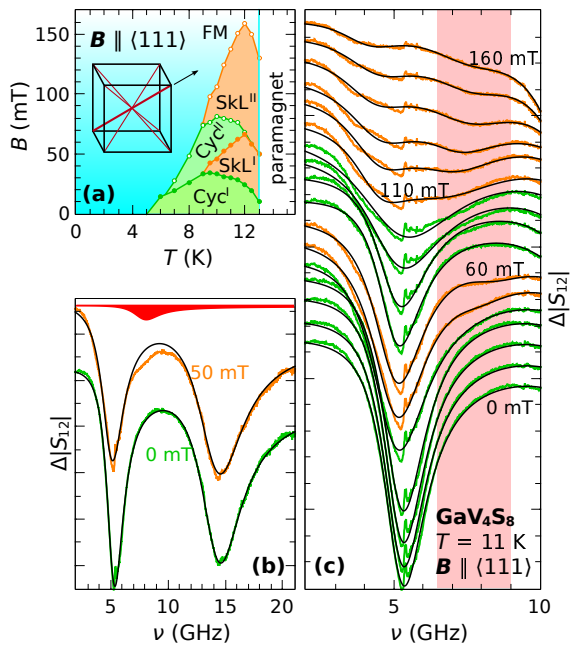}
		\caption{(Color online) (a) Magnetic phase diagram for $B \parallel \langle 111 \rangle$ derived from detailed magnetization measurements \cite{Kezsmarki2015a}. Cycloidal (Cyc$^\text{I}$, Cyc$^\text{II}$) and skyrmion-lattice spin structures (SkL$^\text{I}$, SkL$^\text{II}$) are embedded in a ferromagnetic (FM) phase. The cube and arrow illustrate the orientation of external field with respect to the body diagonals being the possible easy axes. (b) Decomposition of the fit for \unit[50]{mT} with the SkL excitation highlighted in red. For comparison a spectrum with a fit at \unit[0]{mT} is shown. (c) Microwave transmission spectra in GaV$_4$S$_8$ at external magnetic fields $0 \leq \mu_0 H_\text{ext} \leq \unit[160]{mT}$ along $\langle 111 \rangle$ in the frequency range $2 \leq \nu \leq \unit[10]{GHz}$ at \unit[11]{K}. For clarity the spectra have been shifted by a constant value with respect to each other. The field steps are \unit[10]{mT}. Green curves indicate transmission in the cycloidal phases, orange curves in the SkL phases. Black lines represent Dysonian fits (see text). The shaded area indicates the frequency regime where additional excitations emerge in the skyrmion lattice phases.}
		\label{KaskadePD}
	\end{figure}

	Besides the Bloch-type skyrmions reported in the aforementioned works, a N\'eel-type SkL has recently been discovered in GaV$_4$S$_8$~\cite{Kezsmarki2015a}, where the spins rotate radially towards the vortex core. In this semiconductor characterized by V$_4$S$_4$ clusters with spin $S = \frac 1 2$ \cite{Nakamura2005}, a structural Jahn-Teller transition \cite{Pocha2000} at \unit[44]{K} is followed by the onset of magnetic order at the Curie temperature $T_\mathrm C = \unit[13]{K}$. At the structural transition the lattice is stretched along one of the four body diagonals, resulting in a strongly anisotropic easy-axis magnet. The magnetic multi-domain structure strongly depends on the orientation and strength of the applied magnetic field and gives rise to complex magnetic phase diagrams [see Figures~\ref{PDe}(a), \ref{PDe}(c) and \ref{KaskadePD}(a)] including cycloidal (Cyc), SkL, and ferromagnetic (FM) regions. Specifically, the skyrmions do not follow the external magnetic field but are confined to the magnetic easy axes. The phases have been interpreted in terms of a competition of symmetric and anti-symmetric (Dzyaloshinskii-Moriya, DM) anisotropic exchange couplings, as well as the Zeeman interaction \cite{Kezsmarki2015a}. GaV$_4$S$_8$ gained further interest due to its multiferroic properties, with an orbital-order driven polarization below the Jahn-Teller transition \cite{Wang2015} and spin-driven excess polarizations in all magnetic phases \cite{Ruff2015}. The N\'eel-type skyrmions in this compound are dressed with ferroelectric polarization, with a doughnut-shaped ring of polarization around the vortex cores \cite{Ruff2015}. The magnetoelectric effect of these skyrmions is about two orders of magnitude stronger than previously observed for Bloch-type skyrmions in Cu$_2$OSeO$_3$.
	
	Broadband spectroscopy is now timely to explore the dynamics of N\'eel-type SkLs. In this Letter, we present an experimental investigation of the spin dynamics in single crystals of GaV$_4$S$_8$. Performing measurements with the magnetic field applied along the different crystallographic axes $\langle 100 \rangle$, $\langle 110 \rangle$, and $\langle 111 \rangle$, we find a series of characteristic resonances in all three magnetic phases. The spin dynamics of multiferroic compounds are of particular interest as they offer electric-field control of magnetic states, avoiding large power consuming magnetic fields \cite{Bos2008, Gnezdilov2010a, Bordacs2012, Seki2012c, Mochizuki2013, White2014, Kezsmarki2014, Kezsmarki2015b}.

	GaV$_4$S$_8$ singlecrystals were prepared as described in Ref.~\cite{Ruff2015}. Magnetic microwave spectroscopy was performed by inductive coupling of the sample to microwave fields of variable frequency $\nu$ between \unit[10]{MHz} and \unit[26.5]{GHz} above a coplanar waveguide (CPW) \cite{Schwarze2015}. The signal line was $\unit[20]{\mu m}$ wide, smaller than the diameter of the plate-like samples of about 1mm. Such an approach allowed to detect modes of CW, CCW, and BR character in B20 materials \cite{Schwarze2015}.  The external static magnetic field $H_\text{ext} = B / \mu_0$ (where $B$ denotes the magentic flux density and $\mu_0$ the vacuum permeability) was applied perpendicular to the CPW plane. A network analyzer recorded the transmission of the microwave through the CPW in terms of the amplitude of a scattering parameter $|S_{12}|$. The difference technique, subtracting the spectrum $|S_{12}^\text{ref} (\nu)|$ taken at \unit[2]{T} from the recorded spectra $|S_{12} (\nu)|$, reduced field-independent microwave responses in the setup. All spectra shown here are difference spectra $\Delta|S_{12} (\nu)|$.

	Fig.~\ref{PDe} compares transmission spectra obtained at \unit[11]{K} for $B$ applied along $\langle 100 \rangle$ (b) and $\langle 110 \rangle$ (d) directions. For $B = \unit[0]{T}$, i.e., in the cycloidal phase, we observe two well separated absorptions, close to \unit[5]{GHz} and \unit[15]{GHz}. Such a distinct splitting at low fields is characteristic of complex antiferromagnets, as has been studied, e.g., in  doped manganites \cite{Ivannikov2002} where it arises from a competition of DM interaction and crystal field. Interestingly, the existence of well separated low- and high-frequency modes distinguishes GaV$_4$S$_8$ from so-far investigated canonical skyrmion host magnets \cite{Onose2012, Schwarze2015}. The quantitative analysis of the spectra for the cubic $\langle 100 \rangle$ axis is shown in Fig.~\ref{PDe}(b). The data are reasonably described by two Dysonian-shaped lines \cite{Dyson1955}
	\begin{displaymath}
		D (\nu) = \frac{A[\Delta \nu + \alpha (\nu - \nu_0)]}{(\nu - \nu_0)^2 + \Delta \nu^2},
	\end{displaymath}
	superimposed on a weak linear background. Here $A$ is the amplitude, $\nu_0$ the resonance frequency, $\Delta \nu$ the line width, and $\alpha$ the dispersion to absorption ratio. The asymmetry parameter $\alpha$ usually takes into account finite conductivity contributions, but also accounts for phase shifts due to impedance mismatches \cite{Barnes1981}. We find $\alpha \approx 0.45$ to be constant in different fields and for different modes, supporting the latter assumption.
	
	We now discuss the spectra obtained at a finite $B$ of a few \unit[10]{mT}. First we focus on $B \parallel \langle 100 \rangle$ [Fig.~\ref{PDe}(b)] where $B$ encloses the same angle of 55$^\circ$ with all the four possible domains having their easy axes along the $\langle 111 \rangle$ directions. For this field configuration, the magnetic phase diagram is less complex [Fig.~\ref{PDe}(a)] compared to the other two investigated directions [Figs.~\ref{PDe}(c) and \ref{KaskadePD}(a)]. At \unit[80]{mT}, in the SkL phase [see Figs.~\ref{PDe}(a) and \ref{PDe}(b)], we find the low-frequency mode shifted to \unit[4]{GHz}. At the same time, a shoulder is observed indicating an additional resonance at \unit[7.5]{GHz} (highlighted in red color). In the spectrum at \unit[80]{mT}, the resonance near \unit[15]{GHz} is broader compared to $B = \unit[0]{T}$. In Figs.~\ref{PDe}(c) and \ref{PDe}(d), where $B \parallel \langle 110 \rangle$, two domains have their easy axis perpendicular to $B$ and the other two experience an angle of 35$^\circ$ with $B$. In this case, the pronounced low-frequency resonance is  close to \unit[5]{GHz} at \unit[70]{mT}. We again resolve a shoulder, now at \unit[8.5]{GHz}. The broadened high-frequency resonance resides at a frequency slightly smaller than \unit[15]{GHz}. A more detailed spectral evolution is presented for $B \parallel \langle 111 \rangle$.
	
	Fig.~\ref{KaskadePD}(b) compares spectra at \unit[0]{mT} (Cyc$^\text{I}$) and \unit[50]{mT} (SkL$^\text{I}$). The low- and high-frequency modes are found near 5 and \unit[15]{GHz}, respectively, in both cases. At \unit[50]{mT}, the shoulder attributed to the SkL appears at a frequency of 8 GHz. The solid lines, which are fits of the two main modes, clearly document the missing absorption from the SkL indicated in red. Fig.~\ref{KaskadePD}(c) shows the evolution of the modes in the low-frequency region ($2 \leq \nu \leq \unit[10]{GHz}$) between 0 and \unit[160]{mT}. The small peak between 5 and \unit[6]{GHz} is a field-independent artefact. The intensity of the main absorption line strongly decreases with increasing magnetic field and reveals only a minor shift in frequency. The additional mode close to \unit[8]{GHz} is clearly visible from 50 to \unit[60]{mT} (SkL$^\text{I}$) and from 110 to \unit[160]{mT} (SkL$^\text{II}$). This alternating pattern illustrated by green and orange spectra once more corroborates that the presence of three modes is an intrinsic feature of the SkL phases.
	\begin{figure*}
		\includegraphics[width=12.9cm]{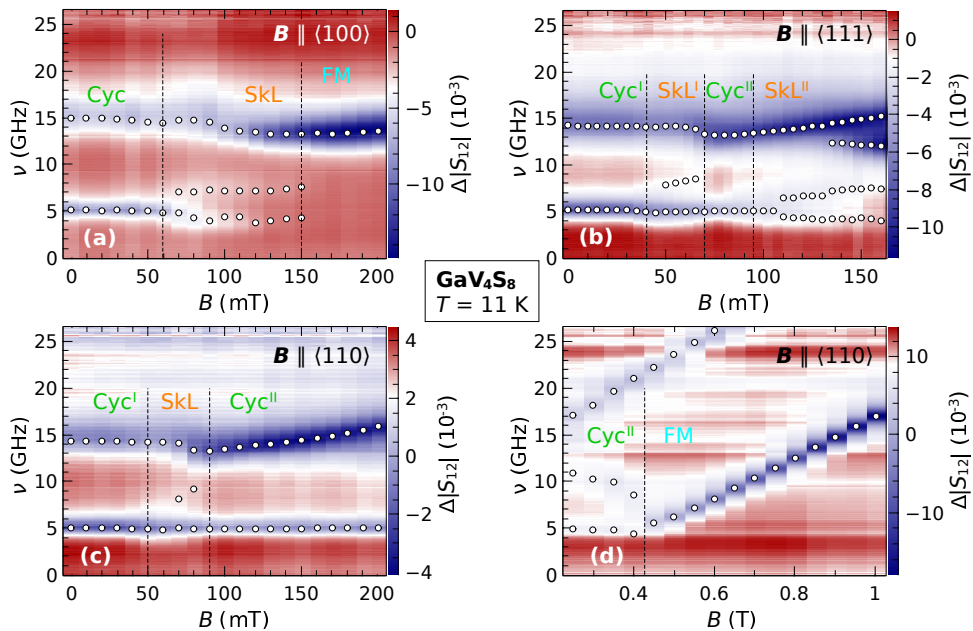}
		\caption{(Color online) Color-coded plots of the transmission spectra $\Delta|S_{12}|$ in the frequency vs.~magnetic field plane for $T = \unit[11]{K}$. Open circles are resonance frequencies from the Dysonian fits (see text). Dashed lines indicate transitions between cycloidal (Cyc), skyrmion lattice (SkL), and ferromagnetic (FM) phases. The field directions are (a) $B \parallel \langle 100 \rangle$ in the range $B \leq \unit[200]{mT}$, (b) $B \parallel \langle 111 \rangle$ in the range $B \leq \unit[160]{mT}$, (c) and (d) $B \parallel \langle 110 \rangle$ in the ranges $B \leq \unit[200]{mT}$ and $0.25 \leq B \leq \unit[1]{T}$, respectively. Along $\langle 111 \rangle$ (b) the FMR splits into two modes (see text).}
		\label{CPlot}
	\end{figure*}
	
	The detailed field dependencies of the excitation spectra along the three principal cubic axes are summarized in Fig.~\ref{CPlot} in color-coded plots. The symbols indicate resonance frequencies extracted from fitting Dysonian-shaped lines to the spectra. In Fig.~\ref{CPlot}a the field $B$ is applied along the $\langle 100 \rangle$ direction. As for this orientation all four rhombohedral domains span the same angle of 55$^\circ$ with $B$ we encounter only two phase boundaries when the SkL and FM phases are entered (vertical dashed lines). At \unit[0]{mT} we observe two excitations close to 5 and \unit[15]{GHz} representative for the cycloidal phase of GaV$_4$S$_8$. In the SkL phase, there are three modes due to the splitting of the low-frequency mode. The lower and upper branches slightly decrease in $\nu$ with increasing $B$ ($\mathrm d \nu / \mathrm d B < 0$). The intermediate resonance residing around \unit[7.5]{GHz}, however, follows the opposite behavior, i.e., $\mathrm d \nu / \mathrm d B > 0$. The upper branch is the most prominent one. At about \unit[90]{mT} this high-frequency mode is relatively broad suggesting a further weak excitation near \unit[17]{GHz} in the SkL. On approaching the FM phase boundary the two low-frequency SkL resonances fade out at about \unit[150]{mT}, while the upper mode strengthens and transforms into a single remaining branch that we attribute to the ferromagnetic resonance. Note that the phase boundaries indicated by dashed lines are somewhat higher than those obtained from the magnetization measurements in Ref.~\cite{Kezsmarki2015a}. The samples used in the microwave experiment are thin plates magnetized perpendicularly to the plane. Due to the demagnetization effect ($N_z = 1$ in the limit of a very thin plate) the internal field $\mathbf{H}_\text{int} = \mathbf{B} / \mu_0 - \overleftrightarrow{N} \mathbf{M}$ \cite{Gurevich1996} is reduced compared to the cube-like crystals ($N_z \approx 1/3$) used in Ref.~\cite{Kezsmarki2015a}. Considering this, our phase boundaries are in fair agreement with the phase diagram reported earlier.
	
	For $B \parallel \langle 111 \rangle$ [Fig.~\ref{CPlot}(b)], the branches show a more complex behavior, as we cross twice the Cyc and SkL phases with increasing $B$. For the domains with the easy axes being parallel to $B$ the phase SkL$^\text{I}$ is entered already at $B = \unit[40]{mT}$. Here the intermediate branch obeys a steeper slope $\mathrm d \nu / \mathrm d B$ compared to Fig.~\ref{CPlot}(a), suggesting that both the applied and the anisotropy field add to the internal field entering the equation of motion \cite{Gurevich1996}. When entering the phase SkL$^\text{II}$ the slope $\mathrm d \nu / \mathrm d B$ of the intermediate branch is smaller. We attribute this to the fact that the relevant domains contributing to the spin resonance have their easy axes tilted by 71$^\circ$ with respect to $B$. In the field regime SkL$^\text{II}$ the lowest branch exhibits $\mathrm d \nu / \mathrm d B < 0$ similar to the SkL phase in Fig.~\ref{CPlot}(a). In SkL$^\text{I}$ of Fig.~\ref{CPlot}(b) we do not resolve this characteristic behavior, most likely due to the superimposed resonance from the domains that still exist in the cycloid state at these fields. At $B > \unit[130]{mT}$, the uppermost branch consists of two resonances. We attribute the upper and lower one to the ferromagnetic resonance (FMR) in domains with the magnetic easy axis aligned with or tilted away from $B$, respectively. 
	
	In Fig.~\ref{CPlot}(c) we display the low-field regime for $B \parallel \langle 110 \rangle$. Here the sample enters the SkL phase for the domains with their easy axes tilted by 35$^\circ$ with respect to $B$. For the domains where these axes are at 90$^\circ$ the applied field cannot induce the SkL phase. Still, the low-frequency mode of the cycloidal phase superimposes on the lowest branch in the SkL phase, similar to SkL$^\text{I}$ in Fig.~\ref{CPlot}(b). This resonance attributed to Cyc$^\text{II}$ exists up to $\unit[0.45]{mT}$ as seen in the high-field data shown in Fig.~\ref{CPlot}(d). This value can be identified as a spin-flop transition of the domains with easy axes perpendicular to the field, regarding the full frequency-field diagram: the upper branch is clearly subdivided into the FMR of the 35$^\circ$-domains that increases linearly with $B$ and the lower, downturning branch attributed to the perpendicular domains. This second branch changes into a second linearly increasing resonance mode above the spin-flop transition at \unit[0.45]{mT}, where the \unit[5]{GHz}-mode disappears.
	
	We now compare our results with the previously reported universal behavior for Bloch-type SkLs \cite{Schwarze2015}. The first significant difference is that in GaV$_4$S$_8$ we observe two modes in the cycloidal phase while in the Bloch-type systems only one mode exists in the helical as well as in the conical phases. This is attributed to the strong magnetic anisotropy in presence of the DM interaction in GaV$_4$S$_8$. In the SkL phase, either of Bloch or of N\'eel-type, three eigenmodes are obtained, indexed in the Bloch-type SkL as CW, CCW, and BR. There the CCW mode (positive slope, $\mathrm d \nu / \mathrm d B > 0$) is lowest in frequency followed by the BR (negative slope) and CW (rather field-independent) modes. In GaV$_4$S$_8$ we find significant positive slope always for the second mode, while the low-frequency mode exhibits a weak negative slope [see Figs.~\ref{CPlot}(a)--\ref{CPlot}(c)] and, hence, we assign the lowest mode as BR mode. This interchange of CCW and BR is attributed to the uniaxial magnetic anisotropy along $\langle 111 \rangle$ that governs the orientation of the skyrmions. For the BR mode, spins that are orthogonal to the easy axis exhibit large precessional amplitude which leads to a reduced effective field in the Landau-Lifshitz equation and therefore shifts the eigenfrequency to small values \cite{Gurevich1996}. Bloch-type skyrmions lack such a contribution to the effective field explaining the different sequence of branches. Now, the intermediate branch with positive slope can be assigned as CCW mode which finally evolves into the FMR, both with the same sense of gyration. At first sight, this is not obvious from Fig.~\ref{CPlot}, but it is indicated by the strong frequency increase of the CCW mode as function of external field in the SKL$^\text{I}$ phases [Figs.~\ref{CPlot}(b)--\ref{CPlot}(c)]. Note that a continuous evolution of the CCW branch into the FMR might not be observed, because the transition from the SkL phase into the FM state is of first order. Returning to the SkL phase, the upper branch can be identified with the CW mode. The CW mode seems to be asymmetric [see, e.g., Fig.~\ref{CPlot}(a)], pointing to an additional mode at even higher frequency, which can be assigned to higher harmonics of the spin-precessional motion.
	
	In summary, after recent reports on magnetic excitations in Bloch-type skyrmion lattices, this work provides a detailed magnetic resonance investigation in a N\'eel-type skyrmion lattice, realized in bulk GaV$_4$S$_8$. Starting from low magnetic fields, in the cycloidal phase, due to competing anisotropic interactions, two resonance modes are observed. Three excitations evolve in the following skyrmion lattice phase which, from lowest to highest frequency, can be ascribed to breathing, counterclockwise, and clockwise modes, with their sequence differing from Bloch-type SkLs. Finally, the CCW mode of the SkL evolves into the ferromagnetic resonance in the field-induced ferromagnetic phase.

	This work was supported by the Deutsche Forschungsgemeinschaft (DFG) via the Transregional Collaborative Research Center TRR 80: From Electronic Correlations to Functionality (Augsburg, Munich, Stuttgart) and by the Hungarian research funds OTKA K 107228, OTKA K 108918, OTKA PD 111756, and Bolyai 00565/14/11. The authors wish to thank D.~Vieweg for performing magnetometry measurements.

\end{document}